\documentstyle[epsfig, subfigure, lscape]{mn}

\title{Particle reacceleration in the Coma cluster:  
       radio properties and hard X--ray emission}
\author[G. Brunetti et al.]
       {G. Brunetti,$^{1,2}$ 
       G. Setti,$^{1,2}$
       L. Feretti,$^{2}$ 
       and G. Giovannini$^{3,2}$\\  
       $^1$ Dipartimento di Astronomia, Universit\'a di Bologna, via Ranzani 1,
       I--40127 Bologna, Italy\\  
       $^2$Istituto di Radioastronomia del CNR, via Gobetti 101,
       I--40129 Bologna, Italy\\
       $^3$Dipartimento di Fisica, Universit\'a 
       di Bologna, via Berti--Pichat 6/2, 
       I--40127 Bologna, Italy\\
}

\begin{document}
\maketitle

\begin{abstract}

The radio spectral index map of the Coma halo 
shows a progressive steepening of the spectral index
with increasing radius.
Such a steepening cannot be simply justified by
models involving continuous injection of fresh particles
in the Coma halo or by models involving diffusion of 
fresh electrons from the central regions.

We propose a {\it two phase} model 
in which the relativistic 
electrons injected in the Coma cluster by some 
processes (starbursts, AGNs, shocks, turbulence)
during a {\it first phase} in the past 
are systematically reaccelerated during a 
{\it second phase}
for a relatively long time
($\sim$ 1 Gyr) up to the present time.
We show that for reacceleration 
time scales of $\sim 0.1$ Gyr this hypothesis 
can well account 
for the radio properties of Coma C. 
For the same range of parameters which explain Coma C
we have calculated  
the expected fluxes from the
inverse Compton 
scattering of the 
CMB photons
finding that the hard X--ray tail discovered 
by BeppoSAX may be accounted for by the stronger
reacceleration allowed by the model.

The possibility of extending the main model
assumptions and findings to the case of the
other radio haloes is also discussed,  
the basic predictions being consistent
with the observations.

\end{abstract}

\begin{keywords}
acceleration of particles - radiation mechanisms: non--thermal -
galaxies: clusters: general - galaxies: clusters: individual: Coma -
radio continuum: general - X--rays: general
\end{keywords}

\section{Introduction}

The intracluster medium (ICM) consists not only
of the hot gas emitting thermal X--rays,
but also of magnetic fields and relativistic
particles.
The existence of cluster magnetic fields
and relativistic electrons is directly
demonstrated by the presence of diffuse
non--thermal 
radio emission detected in a number of clusters, 
the so called radio haloes.
Furthermore, Faraday rotation measurements
toward discrete radio sources
in clusters, coupled with the X--ray emission from the 
hot ICM, have indicated the presence
of $\sim \mu G$ magnetic field strengths.

The radiative lifetime 
of the relativistic electrons, mainly due to 
inverse Compton
(IC) losses with the cosmic microwave background (CMB)
photons, is relatively short ($\sim 10^{7-8}$ yrs)
so that the radio haloes suggest the presence of
relatively fresh injected and/or reaccelerated
electrons in the cluster volume (Jaffe 1977; De Young 1992;
Tribble 1993; Brunetti et al. 1999). 
In general,
models which assume a primary origin of the relativistic
electrons require continuous
injection processes and predict synchrotron radio emission 
and IC emission in the EUV and hard X--rays 
(Ensslin et al. 1997, 1999; Sarazin 1999; 
V\"{o}lk \& Atoyan 1999), while
secondary electron models 
skip the problem of the reacceleration
(Dennison 1980; Blasi \& Colafrancesco 1999)
and predict large gamma--ray fluxes
from neutral pion decay which 
could be tested by future gamma--ray missions.

Among halo sources, Coma C in the Coma cluster
is the most famous and well studied
example of diffuse radio emission in cluster
of galaxies and can be used to test the
different models of radio halo formation and
evolution.

At the lowest frequencies the radio emission 
of Coma C extends up to $\sim 30-40$ arcmin 
from the center (e.g. Cordey 1985 at 151 MHz; 
Hanisch \& Erickson 1980 at 43 MHz;
Henning 1989 at 30.9 MHz); 
at higher frequencies sensitive radio images have been
obtained at 327 MHz (Venturi et al. 1990) and 
at 1.38 GHz (Kim et al. 1990).
By applying Gaussian fits it has been found 
that the 327 MHz FWHM  
(28$\times$20 arcmin) is significantly
larger than that at 1.38 GHz 
(18.7$\times$13.7 arcmin), but
smaller than the low frequency size.
The total radio spectrum of the Coma halo is
steep ($\alpha \sim 1.2$) and   
the 327--1380 MHz spectral index map 
shows that the spectrum steepens rapidly with
increasing distance from the center 
(Giovannini et al.1993).
The steepening has been confirmed 
by 1.4 GHz data from the Effelsberg 
single--dish 100 m telescope
(Deiss et al. 1997).

In principle, current models 
invoking a continuous injection of relativistic electrons
may explain the total synchrotron spectrum
of Coma C, but they
fail in reproducing the spectral 
steepening with radius
(Ensslin et al. 1999; 
V\"{o}lk \& Atoyan 1999). 
Here it is important
to realize that 
the diffusion velocity of the relativistic particles
is low in relation to their radiative lifetimes
so that the spectral steepening cannot be related to the 
diffusion of
fast ageing electrons from the central regions into  
the cluster volume (Berezinsky et al. 1997; Sarazin 1999).
As a consequence, unless one is ready
to accept the rather unphysical scenario of an
injection modulated across the cluster volume,
the spectral steepening
must be related to the intrinsic
evolution of the local electron spectrum and to
the radial 
%shape
profile of the cluster magnetic field strength.

Additional information of great importance
in constraining the models of Coma C
can also be obtained 
from observations at higher 
frequencies.
Recently, Fusco--Femiano et al.(1999)
have discovered an hard X--ray tail exceeding the
thermal emission.
Also in this case the existing models
are not able to reconcile an IC origin of the hard X-ray flux
with the estimates 
magnetic field intensities derived from RM measurements, 
the required field being at least one order of 
magnitude weaker than the RM estimates
(e.g. Sarazin 1999, Ensslin et al. 1999 and 
references therein).
This has stimulated alternative proposals
for the origin of the hard X--ray excess 
(Blasi, Olinto, Stebbins 2000; 
Ensslin 2000 and references therein).

In view of these difficulties
we investigate the alternative possibility
and potentiality of a {\it two phase} scenario
for Coma C in which 
relativistic particles injected during
a {\it first phase} are then reaccelerated during a 
{\it second phase}
possibly associated with a recent merger event as
indicated by optical and X-ray observations.
A preliminary, much simplified version of the model
has been presented elsewhere (Brunetti et al.1999). Here
radiation and Coulomb losses and reacceleration efficiency
are fully taken into account as a function of the cluster
radius and the assumption of a constant magnetic field 
throughout the cluster volume is released. A detailed 
description of the time evolution of the electron spectra
is a distinctive feature of our model as compared to other
current models (e.g. Sarazin 1999 and Ensslin 2000 for a review)

The main aims of the present work are to:

i) investigate the effects of energy losses and reacceleration 
gains on
the energy distribution of the relativistic electrons and
find out whether the radio spectral steepening may
significantly constrain the space of parameters
of the physical
processes at work in Coma C;

ii) obtain 
the radial trend of the magnetic field strength  
for the same range of the physical parameters in i)
and compare it with the available information;

iii) 
compare the derived IC emission of Coma C 
%implied by the above model parameters and compare it 
with the hard X-ray flux measured by BeppoSAX 
and find out whether  
the relatively high central values of the magnetic field
indicated by the RM results
are compatible with the IC scenario.

In Section 2 we introduce the model; Section 3 is devoted 
to a general presentation of the time evolution of the 
electron energy distribution subject to reacceleration and
losses; the 
results concerning the radio properties of Coma C are
given in Section 4, while those concerning
the hard X--rays are given in
Section 5. In Section 6 we discuss our findings and
extend the model for Coma C to the case of
the other haloes, while the main conclusions are summarized
in Section 7. In the Appendix we present a 
more detailed discussion of the theoretical aspects
of the model.
$H_0 = 50$ km s$^{-1}$ Mpc$^{-1}$ 
is assumed throughout.

\section{The two phase model}

\subsection{The radio spectral steepening}
\begin{figure}
\resizebox{\hsize}{!}{\includegraphics{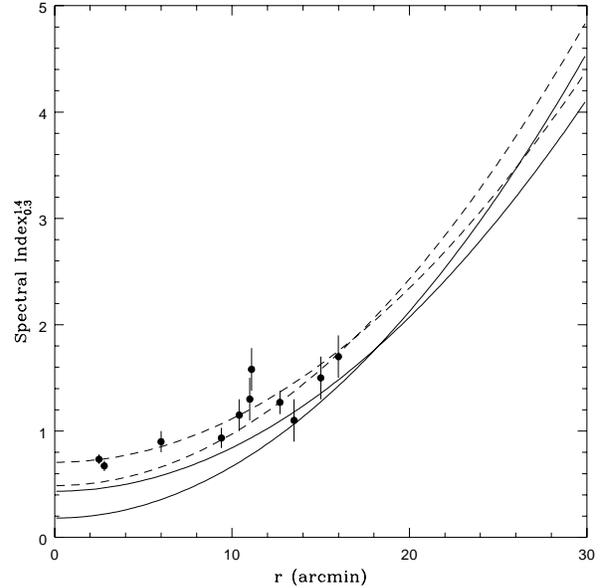}}
\caption[]{The predicted observable (dashed lines)
and intrinsic (solid lines) spectral index 
for the Coma halo
are reported as a function of the distance from the center.
The curves are obtained from Deiss et al.(1997) relationships
based on 327/1380 MHz (bottom dashed and solid lines
at $r$=0)
and 327/1400 MHz (single dish) observations.
The points are taken from three slices 
of the spectral index map (327--1380 MHz) 
of Giovannini et al.(1993)
across the halo center.} 
%(different simbols represent different slices).}
\end{figure}

By assuming a spherical symmetry of the
relevant physical quantities and a Gaussian spatial profile
of the emission coefficient, 
Deiss et al.(1997) have derived relationships between 
the observed and intrinsic spectral index
distributions of the synchrotron emission. 
In the case of the Coma halo they have shown
that the FWHMs and flux densities of the 
radio observations (Venturi et al. 1990; Kim et al.
1990) can be reproduced by a central intrinsic spectral index
$\alpha \sim 0.2-0.4$,
leading to a central observed spectral index  
$\alpha \sim 0.7$,
in agreement with Giovannini et al. (1993) findings.
In Fig.1 we plot the 
observed and intrinsic spectral indices predicted 
from Deiss et al. (1997) relationships based on
the measurements at 327 MHz (Venturi et al.1990), 1.38 GHz
(Kim et al. 1990) and 1.4 GHz (Deiss et al. 1997).
Both the data at 1.38 GHz 
(obtained with a synthesis radio telescope) 
and those at 1.4 GHz
(obtained with a single dish radio telescope),  
when combined with the 327 MHz data,  
show a marked steepening of the
observed spectrum with distance from the center; this
confirms that the  
steepening in the Giovannini et al.(1993) 
spectral index map is not due to flux density losses
at 1.38 GHz but it is real.
The points reported in Fig.1 are taken from 
three slices across the halo center 
of the spectral index map 
(Giovannini et al. 1993).
Although there is some scattering,
the agreement between the observed and derived spectral
distributions is fairly good.

\subsection{Injection Phase}

We investigate the application of 
a {\it two phase} model for Coma C.
During the {\it first phase} 
fresh particles are continuously injected 
in the volume of the Coma cluster 
for a sufficiently long time.
An injection {\it phase} in the past 
can be reasonably related to starburst 
activity, during which violent galactic
winds can accelerate relativistic particles
(Zirakashvili et al. 1996; Ptuskin et al. 1997),   
and/or to the activity of AGNs 
producing a large amount of relativistic particles
(Giovannini et al. 1993; Sarazin 1999). 
In addition, the high temperature of the 
ICM suggests that the ICM  has passed through 
strong shocks, possibly during major mergers events when  
a significant portion of the shock energy can be channeled
into the acceleration of relativistic particles 
(Blandford \& Eichler 1987; Jones \& Ellison 1991). 
Moreover, it is also possible that 
during major mergers relic particles,  
released in the past from starbursts and  
AGN's activity, can be reaccelerated and injected 
at higher energies.

The time evolution of the particle energy distribution
during the injection, $N(\gamma,t)$,  
is obtained by solving the kinetic equation (e.g. Kardashev 1962):

\begin{equation}
{{\partial N(\gamma, \theta, t)}\over
{\partial t}} =
-{{\partial}\over{\partial \gamma}} 
\left[ {{d \gamma}\over{d t}} N(\gamma,\theta, t) \right]
+ Q_{inj}(\gamma,\theta, t)
\label{kinetic}
\end{equation}

\noindent
with 
$Q_{inj}(\gamma, \theta, t)$ the injection function.

\noindent
Following Sarazin (1999; Fig.1)
we assume that the cooling of the relativistic electrons
is dominated
by Coulomb ($\xi$) and radiation($\gamma^2 \beta(z,\theta)$)losses, 
the 
relativistic bremsstrahlung
being negligible, so that

\begin{equation}
{{d\gamma}\over{d t}} =
-\xi - \beta(\theta\,,z) \gamma^2
\label{fase1losses}
\end{equation}

\noindent
The Coulomb losses depend on the ICM density
as (e.g. Sarazin 1999):

\begin{equation}
\xi \simeq 1.2\cdot 10^{-12} n 
\left(1 + {{ln(\gamma/n)}\over{75}} \right)
\sim 1.4 \cdot 10^{-12} n \,\,\,\, s^{-1}
\label{culomblosses}
\end{equation}

\noindent
while the coefficient of the radiative losses
is:

\begin{equation}
\beta(\theta\, ,z)=
1.9 \cdot 10^{-9} 
\left( B^2 sin^2\theta + B_{IC}^2(z) \right)
\,\,\,\,\, s^{-1}
\label{radiativelosses}
\end{equation}

\noindent
where $\theta$ is the pitch angle,
$z$ the redshift and $B_{IC}$  
the equivalent magnetic field
strength of the CMB.

\noindent
If the particles are isotropically injected, 
i.e. $Q_{inj}(\gamma, \theta)= Q_{inj}(\gamma) / (4\pi)$, 
the only dependence of the electron spectrum
on the pitch angle is given
by the synchrotron loss term (Eq.\ref{radiativelosses}).
If the pitch angle scattering is an efficient process then
the electrons are continuously isotropized, 
the term $B^2 sin^2\theta$ should be replaced by
$2/3 B^2$ and the solution is isotropic, i.e.
$N(\gamma, \theta) = N(\gamma) / (4\pi)$.

\noindent
The IC losses depend on the redshift 
as $(1+z)^4$ so that
the cooling time of the relativistic electrons 
in clusters is expected to be dominated by the
IC process at moderately and high redshifts; this produces a
rapid ageing of fresh particles injected at $z > 0.3$
(Sarazin 1999).
Due to the presence of reacceleration that dominates the
time evolution of the electrons 
during the {\it second phase} and since it starts 
less than $\sim$1 Gyr before
the corresponding cluster age, 
in our model the dependence on $z$ of the radiative
losses is not important and has been neglected in the 
calculations.

\noindent
Since the injection phase is assumed to hold 
in the past for a time
considerably longer than the 
the cooling time of the electrons, 
the energy distribution
approaches a stationary spectrum (Sarazin 1999). 
Under stationary conditions 
and assuming a time--independent 
$Q_{inj}= K_e \gamma^{-\delta}$, 
Eqs.(\ref{kinetic}) 
and (\ref{fase1losses}) yield:

\begin{equation}
N(\gamma,\theta)= {{K_e}\over{\beta(\theta) (\delta-1)}}
\gamma^{-\delta+1}
\left( \gamma^2 + {{\xi}\over{ \beta(\theta)}} \right)^{-1}
\label{fase1}
\end{equation}

\noindent
which is $\propto \gamma^{-\delta+1}$ at low energies (where
Coulomb losses dominate) and 
$\propto \gamma^{-\delta-1}$ at higher energies
where radiative losses dominate.
\footnote{we will assume $\delta$=2.5 throughout this
paper}

\subsection{Reacceleration Phase}

When the injection of fresh particles is stopped
or strongly reduced the energy
distribution rapidly steepens and the number of
high energy radio emitting electrons  
rapidly decreases.

In order to avoid the problem of the rapid ageing,  
we investigate the possibility that some 
reacceleration in the cluster volume is present
during a {\it second phase} 
lasting less than about 1 Gyr 
up to the present cluster age.
The presence of reacceleration mechanisms
in cluster of galaxies has been pointed out  
by many authors studying the spectral ageing
in head--tail radio sources 
(e.g. Jaffe \& Perola 1973; Ekers et al. 1978; 
Fanti et al. 1981; Parma et al. 1999).

As a simplification of the scenario, 
we assume that the 
{\it second phase} starts when the efficiency of 
the injection processes is strongly reduced or stopped.
The general case in which the end of the injection 
and the start of the reacceleration do not coincide
is given in Appendix A; the main results 
remain unchanged.
The {\it second phase} may be related to a merger or a 
post--merger scenario 
in which shocks and cluster weather,  
powered by the energy released during a merger,  
can compensate at some level for the 
radiative and Coulomb losses.
A number of authors have found 
evidence for recent merging in the Coma cluster 
(Briel et al. 1992; Vikhlinin et al. 1997). 
Although we discuss specifically 
the case of Coma C, 
the radio haloes are generally found in clusters
with recent merger activity  
(e.g. Feretti \& Giovannini 1996); the general application
of the {\it two phase} model to the radio haloes is discussed
in Sect.7.2.

If the observed radio spectral steepening is related
to the intrinsic evolution of the local electron energy
distribution rather than to an unphysical injection
modulation, 
important constraints on the physical processes at work in
Coma C can be immediately obtained.

Indeed, the spectral steepening implies that
the break of the synchrotron spectrum is   
in the range 0.3--1.4 GHz for a relevant fraction of the 
cluster volume.
By making use of the basic synchrotron relationships 
this immediately yields a typical break energy of the  
electron spectrum at 
$\gamma_b \sim 10^4 (B_{\mu G})^{-1/2}$ and a
radiative lifetime  
$T \sim 5 \cdot 10^{15} ( B_{\mu G})^{1/2} \, s$
for relatively weak magnetic fields 
($B < B_{IC} \sim 3 \mu$G).
Therefore, in order to generate Coma C the reacceleration 
during the {\it second phase} must essentially balance the
radiation losses, i.e. the reacceleration efficiency 
should be 
$\chi \sim 1/T \sim 2 \cdot 10^{-16} / \sqrt{ B_{\mu G} }
\, s^{-1}$. Moreover, a significant flattening of the
relic electron distribution may only be attained for a 
reacceleration time $\tau$ such that electrons of energy 
$\gamma \sim$ 100 are accelerated to $\sim \gamma_{b}$ 
(Section 3 for details and Eq.(\ref{fase2losses})), so that 
$\tau > ln(\gamma_{b}/100) \chi^{-1} \sim$ 0.6--0.7 Gyr.

If the average magnetic field intensity depends on
the radial distance $r$ from the cluster centre, 
it clearly follows that
a reacceleration efficiency
$\chi(r) \propto 1/\sqrt{B(r)}$ would produce
a constant radio spectrum  throughout the cluster.
By allowing for a likely decrease of B with r, it then
follows that a constant reacceleration efficiency 
results in a systematic steepening of the synchrotron spectrum
with $r$ simply because at a given frequency higher energy 
electrons are selected by the lower field intensity.
This steepening effect would be enhanced if the reacceleration
efficiency increases in the regions of stronger fields.
On the other hand, Coulomb losses are particularly severe
in the central region of the cluster leading to a marked
flattening of the electron spectrum at low energies and,
consequently, to a flattening of the synchrotron spectrum
at low frequencies as observed.

The above qualitative description of the effects of systematic
reacceleration and losses on the evolution of the electron
spectrum will be discussed in detail in the following Section 3.
(Readers specifically interested in the application of the 
model to Coma C may skip this Section and go directly 
to Section 4.)

\section{Evolution of the electron and synchrotron spectra
}

In this Section we illustrate the time evolution
of the electron energy distribution and related synchrotron 
spectra for a wide range of the relevant
physical parameters. This provides the basic quantitative
scenario delimiting the region of the parameters' space
which produces a self-consistent picture of the emission 
from Coma C.

\subsection{The electron spectrum}

As a simplification 
we assume that
the particles are systematically reaccelerated
by first order Fermi mechanisms.
The time evolution of the energy of the electrons is: 

\begin{equation}
{{d\gamma}\over{dt}}
=-\beta(\theta) \gamma^2 -\xi+ \chi \, \gamma
- \eta(\gamma) \gamma 
\label{fase2losses}
\end{equation}

\noindent
where
$\chi$ expresses the efficiency of the 
reacceleration 
and $\eta(\gamma)$ the relativistic 
bremsstrahlung
losses given by (e.g. Sarazin 1999)

\begin{equation}
\eta(\gamma)
\simeq
1.5 \cdot 10^{-16} n\, [ ln(\gamma) +0.36] \,\,\,\,\,
s^{-1}
\label{bremss}
\end{equation}

Given the typical 
cluster thermal density $n \sim 10^{-3}cm{-3}$, 
the bremsstrahlung losses
are always $\sim 100$ times lower than the
reacceleration efficiencies adopted in this work
and will be neglected.

With the initial ($t=t_i$) spectrum given by
Eq.(\ref{fase1}) the solution of the kinetic equation
(Eq.\ref{kinetic} with $Q=0$) 
is:

\begin{eqnarray}
N(\gamma,\tau,\theta)= 
{{K_e q(\theta) }\over{\delta-1}}
{{  [1-\tanh^2(\tau\sqrt{q}/2)] \gamma^{-\delta+1} }\over{
[2 \gamma_b(\tau,\theta) \tanh (\tau \sqrt{q} /2)]^{2} \beta^3(\theta)}} 
 \nonumber\\
\cdot
\left(1- {{\gamma}\over{\gamma_b(\tau,\theta)}} \right)^{\delta-3}
\left( 1+ {{\xi \gamma^{-1} -\chi}\over
{\gamma_b(\tau,\theta) \beta(\theta)}} \right)^{-\delta+1} 
\cdot \nonumber\\ 
\left( \gamma^2
\left[ {{1 + (\xi \gamma^{-1} -\chi)(\gamma_b(\tau,\theta)
\beta(\theta))^{-1} }\over{
1- \gamma/\gamma_b(\tau,\theta) }} \right]^2
+ {{\xi}\over{\beta(\theta)}} \right)^{-1}
\label{fase2}
\end{eqnarray}

\noindent
where 
$q(\theta)  = \chi^2-4\xi \beta(\theta)$ 
($q>0$; see Appendix A), 
$\tau = t -t_i$, 
and $\gamma_b$ is the break energy (i.e. the largest 
energy of the electrons) given by  

\begin{equation}
\gamma_b(\tau,\theta)=
{{1}\over{2\beta(\theta)}} \left( \chi
+ {{ \sqrt{q} }\over{ \tanh {{\tau \sqrt{q}}\over{2}} }} 
\right)
\label{gammab}
\end{equation}

\noindent
For $\xi =0$ (no Coulomb losses) one finds
$\gamma_b(\theta) = \chi /\beta(1-e^{-\chi \tau})$
in agreement with Kardashev (1962)
results.

\begin{figure}
\resizebox{\hsize}{!}{\includegraphics{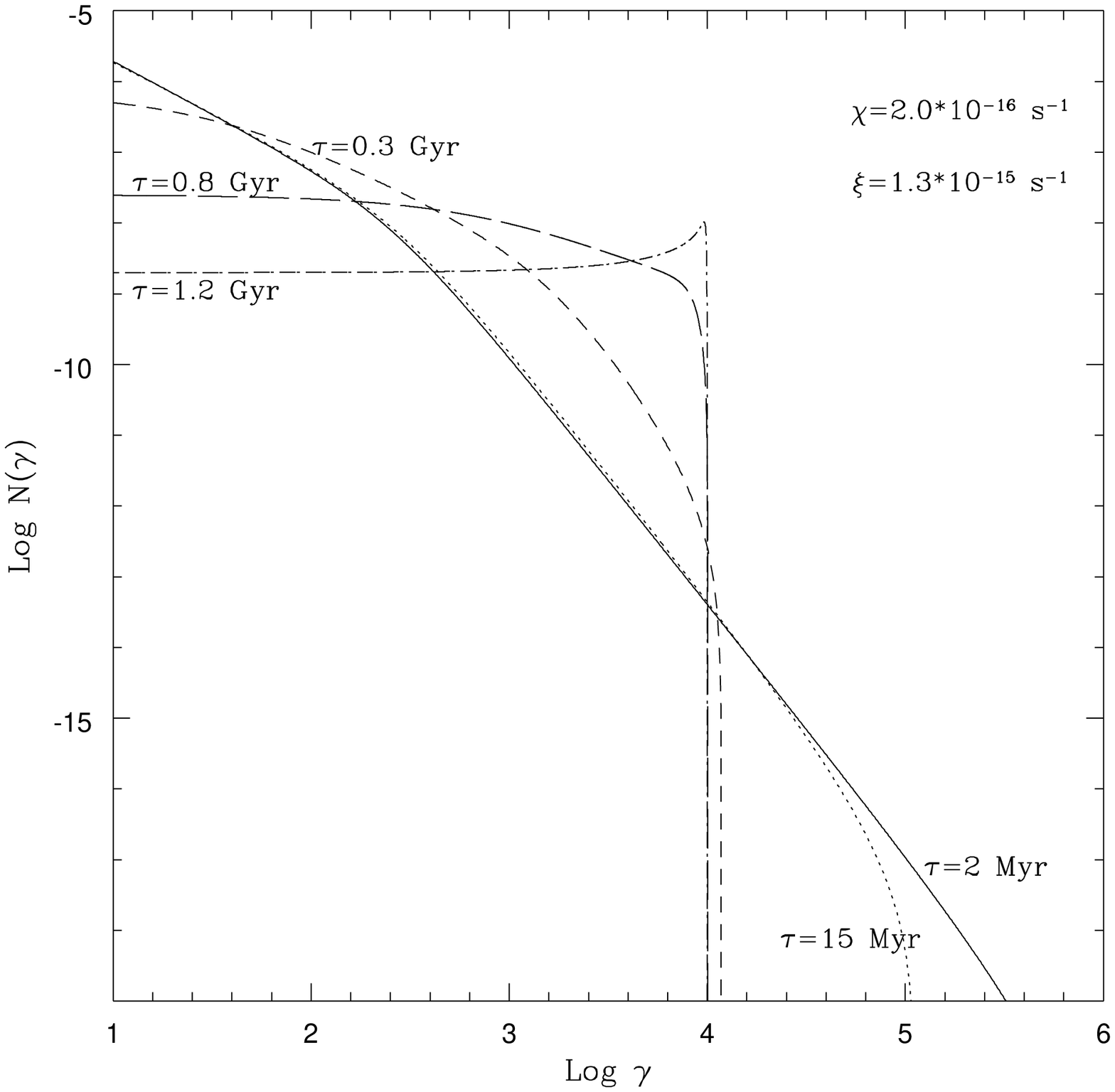}}
\caption[]{
The predicted electron 
energy distribution is shown 
in arbitrary units as a function of 
the reacceleration period 
$\tau$ as indicated in the panel.
The calculations are performed from Eq.(\ref{fase2}) for 
$\chi=2.0 \cdot 10^{-16}$
s$^{-1}$, 
$\xi = 1.3\cdot 10^{-15}$s$^{-1}$ and $B=0.0$.}
\end{figure}
\begin{figure}
\resizebox{\hsize}{!}{\includegraphics{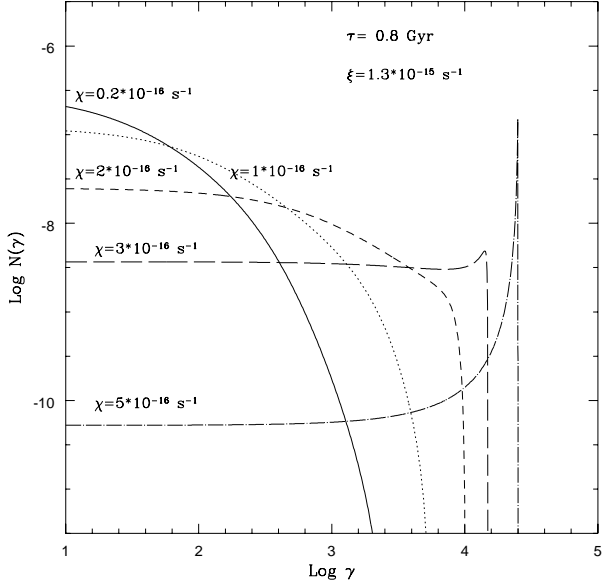}}
\caption[]{The predicted electron 
energy distribution is shown 
in arbitrary units as a function of the 
reacceleration efficiency as indicated in
the panel
The calculations are performed from Eq.(\ref{fase2})
for $\tau=0.8$ Gyr, 
$\xi = 1.3\cdot 10^{-15}$s$^{-1}$ and $B=0.0$.}
\end{figure}

An example of the evolution with time 
of the energy distribution 
of the electrons is
represented in Fig.2 for $\chi=2.0\cdot 10^{-16}$ s$^{-1}$
(i.e. a reacceleration time $\simeq 1.6 \times 10^{8}$ yr)
and $\xi=1.3\cdot 10^{-15}$ s$^{-1}$
(i.e. a thermal gas density 
$n \sim 10^{-3}$ cm$^{-3}$).
  
Due to the radiative losses  
the high energy part of the distribution  evolves 
rapidly with time. 
When the break energy $\gamma_b$ reaches the asymptotic
value $(\chi+\sqrt{-q})/2\beta$ the radiation losses
are balanced by the reacceleration and 
the break energy is frozen;
when $\tau$ becomes
sufficiently larger than the reacceleration time
the combined effects of Coulomb losses and 
systematic reacceleration cause a depletion
of the low energy side of the electron energy distribution
which rapidly flattens.

In Fig.3 we report the electron energy distribution 
after a representative time interval $\tau = 0.8$ Gyr 
for different reacceleration efficiencies.
As expected, one finds that stronger 
reaccelerations yield higher energy breaks and flatter
energy distributions.

\begin{figure}
\resizebox{\hsize}{!}{\includegraphics{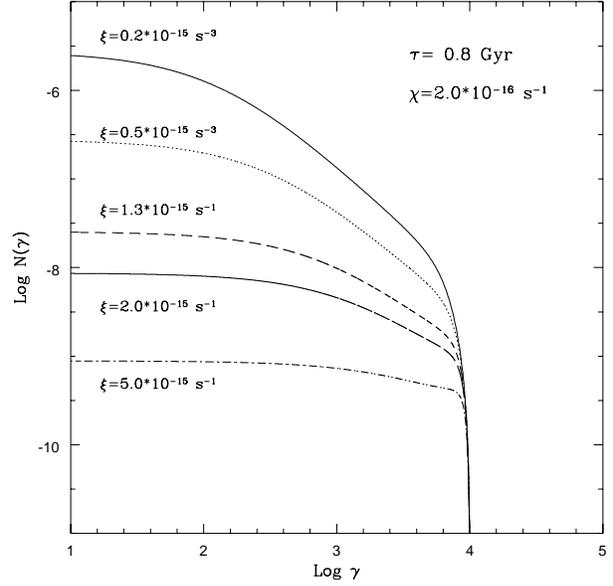}}
\caption[]{The predicted electron 
energy distribution is shown in arbitrary 
units as a function of 
the Coulomb losses as indicated in the panel.
The calculations are performed from Eq.(\ref{fase2})
for $\chi=2.0\cdot 10^{-16}$
s$^{-1}$, $\tau=0.8$ Gyr and $B=0.0$.}
\end{figure}
\begin{figure}
\resizebox{\hsize}{!}{\includegraphics{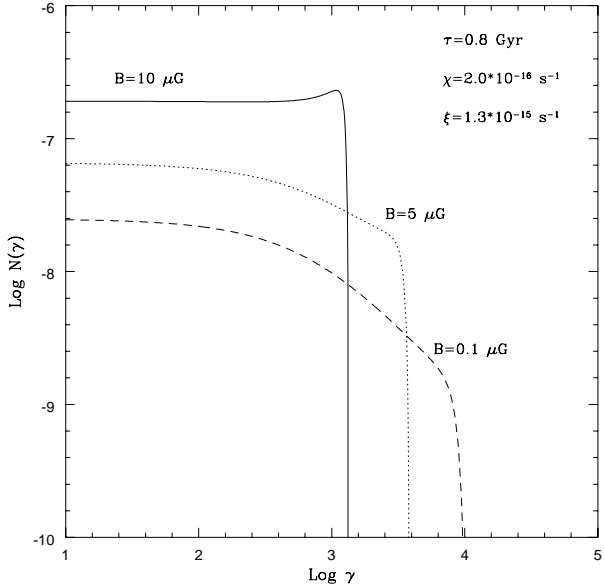}}
\caption[]{The predicted electron 
energy distribution is shown in arbitrary 
units as a function of 
the magnetic field strength $B$ as indicated
in the panel
The calculations are performed from Eq.(\ref{fase2})
for $\chi=2.0\cdot 10^{-16}$ s$^{-1}$, 
$\xi=1.3\cdot 10^{-15}$ s$^{-1}$,
$\theta=55^o$ and $\tau=0.8$ Gyr.}
\end{figure}

The effect on the particle
energy distribution due to the increase 
of the Coulomb losses is illustrated 
in Fig.4 for a given reacceleration. 
By increasing the ICM density 
the Coulomb break in the initial
spectrum is positioned at larger $\gamma$'s, 
causing a larger flattening
of the electron spectrum at lower energies
(Eq.\ref{fase1}), so that the reaccelerated 
spectrum is also flatter. 
At this point it should be noticed that,  
since the ICM density falls down with
increasing distance from the center, also  
the electron energy distribution is expected
to be flatter in the central regions of the clusters.

The effect of increasing the radiative losses 
is illustrated in
Fig.5 for a given reacceleration.
As expected, the balance between losses and
reacceleration is obtained at lower energies with
increasing $B$.
Furthermore, with increasing $B$  
the break energy $\gamma_b$ becomes closer to
the energy at which the reacceleration balances the Coulomb
losses and the energy distribution slightly flattens.

All these findings hold in general:
in the case of systematic
reacceleration of relativistic electrons diffuse 
in a thermal medium (as in clusters of
galaxies) after a time sufficiently large,  
which depends on 
the reacceleration and/or losses,  
the electron energy distribution is expected to 
be flat or slightly inverted 
before a sharp cut--off.

\subsection{The synchrotron spectrum}

\begin{figure}
\resizebox{\hsize}{!}{\includegraphics{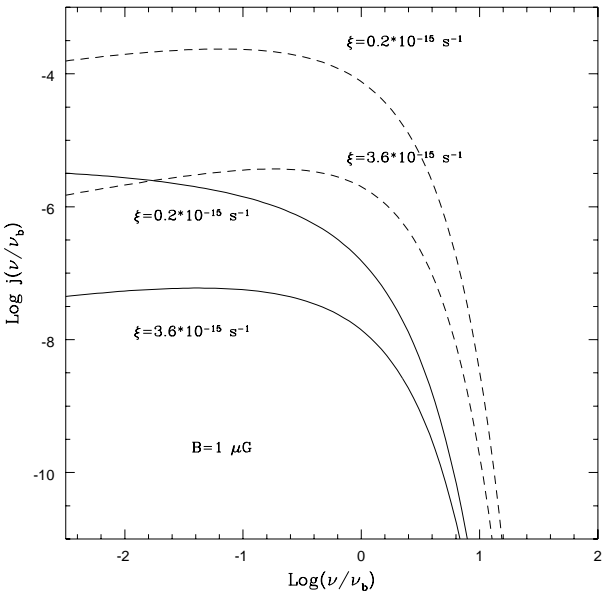}}
\caption[]{The predicted synchrotron
spectrum is shown in arbitrary units 
for two well separated values of the Coulomb losses
as indicated in the panel.
The frequency scale is in unit of the break frequency
$\nu_b$.
We assume two reacceleration efficiencies:
$\chi = 2.5 \cdot 10^{-16}$ s$^{-1}$ (solid lines)
and $\chi =4.2 \cdot 10^{-16}$ s$^{-1}$ (dashed lines).
The calculations are performed for 
a reacceleration period $\tau$=0.8 Gyr and 
$B=1.0 \mu G$ (typical for the Coma center).}
\end{figure}

In general the synchrotron emissivity is obtained by 
integrating the synchrotron Kernel over the
electron energy and angular distributions 
(cf. Westfold 1959; Pacholczyk 1970).
We assume that the electron momenta are isotropically
distributed and that the fields are tangled and of constant
intensity on a sufficiently small scale.
These assumptions allow to consider the synchrotron
emissivity (averaged over a sufficiently large volume) 
isotropic. 
By omitting for simplicity of notation the dependences of
the functions, 
the emissivity per unit solid angle is :

\begin{eqnarray}
j\left({{\nu}\over{\nu_b}},\tau\right)= 
{{K_e }\over{\delta-1}}
{{ \sqrt{3} }\over{4 }}
{{e^3}\over{ mc^2}} B
\int_0^{\pi/2} d\theta sin^2 \theta 
\phi(\theta) \cdot \nonumber\\
\int_0^1 du
F\left({{\nu/\nu_b}
\over{u^2 sin\theta}}\right) 
( 1 - u )^{\delta -3}
u^{-\delta +1} \cdot
\nonumber\\
\left\{ 1+ 
{{ \xi u^{-1} -\chi \gamma_b(\theta)}\over
{\beta(\theta) \gamma_b^2(\theta)} } 
\right\}^{-\delta+1} \cdot 
\nonumber\\
\left\{ u^2 [ {{ 1 +(\xi u^{-1} -\chi \gamma_b(\theta)) 
/(\gamma_b^2(\theta) \beta(\theta))}\over{ 1- u }} ]^2 
+ {{\xi}\over{\beta(\theta) \gamma_b^2(\theta)}} 
\right\}^{-1} 
\label{sincroemiss}
\end{eqnarray}

\noindent
where the electron Kernel 

\begin{equation}
F(y)= y \int_y^{\infty} K_{ {5\over 3} } (v) dv
\end{equation}

\noindent
$ K_{ {5\over 3} }$ being the 
 $2/3$ order modified {\it Bessel}
function, 

\begin{equation}
\phi(\theta)= q(\theta) {{ 
 1 - \tanh^2(\tau \sqrt{q}/2\,)  }\over
{ [\tanh (\tau \sqrt{q}/2\,) ]^{2} }} 
\gamma_b(\theta,\tau)^{-(\delta+2)}
\beta^{-3}(\theta)
\label{phisyn}
\end{equation}

\begin{equation}
u (\theta) =  {{\gamma}\over{\gamma_b(\theta.\tau)}}
= C  
{{ B^2 sin^2\theta +B_{IC}^2 }\over
{ \chi + 
\sqrt{q(\theta)}/ \tanh( {{\tau \sqrt{q}}\over{2}}) }}
\gamma 
\label{gammab2}
\end{equation}

\noindent
$C \simeq 3\cdot 10^{-9}$ cgs and 
$\nu_b$ the critical 
frequency calculated for $\gamma=\gamma_b$
($\theta=90^o$).

As it is well known, 
the shape of the synchrotron spectrum depends
on the electron distribution in the
3--D momentum space; this is a function of the
combined losses and reaccelerations (Eq.\ref{fase2}).
The IC losses are constant over
the cluster volume, but both the magnetic field $B$ 
and the Coulomb losses are expected to 
vary with the distance $r$ from the center.
Thus, the relativistic electron energy
distribution (Eq.\ref{fase2}), 
the break energy and the importance of the
synchrotron losses
with respect to the IC losses  
depend on $r$.
As a consequence the shape of the synchrotron
spectrum (Eqs.\ref{sincroemiss}, \ref{phisyn})  
depends on $r$ and it is expected to be 
intermediate between those calculated 
in the case of pure synchrotron losses
(Kardashev 1962; Pacholczyk 1970; Komissarov 
\& Gubanov 1994 : KP--model) and of 
pure IC losses (Jaffe \& Perola 1973;
Komissarov \& Gubanov 1994: JP--model).

As discussed in the preceeding Section 
increasing the Coulomb losses and/or 
the systematic reacceleration yields, in general,  
a flattening in the energy distribution of the 
relativistic electrons.
The effects on the synchrotron
spectrum are illustrated in Fig.6
for given $\tau$ and $B$.
As expected, 
the synchrotron spectrum 
below the break frequency 
becomes flatter with increasing the 
ICM density 
and/or the systematic reacceleration.
It should also be noticed that the synchrotron
specra are not sensitive to the presence of spectral
spikes in the electron energy distribution (Fig.3).
If stochastic Fermi mechanisms 
provide a substantial contribution 
to the reacceleration of the electrons,  
the spectral spikes are expected to be
smoothed out, but our results remain unchanged.

\section{Model results for Coma C}

As shown in the previous Section 
a natural consequence of the combined reacceleration
and Compton losses is the flattening of the
electron spectrum and of the related synchrotron 
emission in the central region of a cluster,
in qualitative agreement with what is actually found
in Coma C.
In this Section
the physical parameters 
will be constrained in order to reproduce 
the observed spectral steepening and 
total radio spectrum.

\subsection{Basic assumptions}

Following Deiss et al.(1997) we approximate the 
brightness distribution at each frequency 
with a symmetric Gaussian of  
FWHM $\Delta(\nu)$ equal to the geometric mean of 
the smaller and larger 
FWHM measured at each frequency.
The very rapid decrease of the radio brightness given by the
Gaussian approximation 
cannot well constrain the synchrotron electron
distribution at large distances from the cluster 
center where the brightness fades into the noise. 
Thus, in the model
computation, the tail of this distribution should be allowed
to vary within the observational uncertainties.
To this end we model the radio brightness 
distribution by adding to 
the Gaussian from the fits at each observed frequency 
a symmetric Gaussian of larger FWHM $\Delta_l(\nu)$ 
so that the observed emissivity is represented by:

\begin{equation}
j^{obs}(r,\nu)= A(\nu) \cdot \left\{
e^{ -({r\over{0.6\,\Delta(\nu)}})^2 }
+ f e^{ -({r\over{0.6\,\Delta_l(\nu)}})^2 }
\right\}
\label{radiobr}
\end{equation}

\noindent
$A(\nu)$ giving the normalization of the radio
brightness distribution and $f$, a free parameter  
(from the radio maps we find $f\leq 0.06$).

\noindent
As shown in the preceding Section 
the Coulomb losses play an important role in
shaping the electron energy distribution 
and the associated synchrotron spectrum. 
By adopting the $\beta$--model of  
Briel et al.(1992) the Coulomb losses are :

\begin{equation}
\xi(r)=
4.2 \cdot 10^{-15}\, 
\left( 1 + ( {{r}\over{R_C}} )^2
\right)^{-3\beta/2}
\label{thermdis}
\end{equation}

\noindent
where 
$R_C$=10.5 arcmin is the core radius of the ICM
and $\beta=0.75$ (the central density is 
$n_0 \simeq 3 \cdot 10^{-3}$ cm$^{-3}$). 

We assume that the efficiency of the reacceleration 
can be parameterized  as  
the sum of a constant large scale component 
and a small scale component.
Indeed, turbulent gas motion originating
from the massive galaxies orbiting in the cluster   
can amplify the efficiency of the large 
scale reacceleration  
within the cluster core (Deiss \& Just 1996).
As a possible choice, we assume that the small scale 
reacceleration is proportional to the inverse of the 
typical distance between galaxies.
By adopting the observed galaxy distribution in the Coma
cluster (Girardi et al. 1995), 
we parameterize the reacceleration as : 

\begin{equation}
\chi(r)= \chi_{LS} + 
\chi_{Go} \left( 1+ ( {r\over
{R_{G}}} )^2 \right)^{-\alpha_G/3}
\label{alpha}
\end{equation}

\noindent
$R_{G}$=5.5 arcmin being the optical core radius and
$\alpha_G$=0.8.

It should be noticed, however, that 
since the radio brightness and the
spectral properties of 
the model are obtained by integrating
the emissivities over the cluster volume, the main 
model results do not strongly depend on 
the precise choice 
of the functional dependence in Eq.(\ref{alpha}).

In the sequel we adopt a reacceleration time interval
$\tau = 0.8 Gyr$, but we anticipate that the essential features
of the model are not modified as long as $\tau$ is of the order
of 1 Gyr.

At last we assume complete
reisotropization of the electron momenta, 
not only for simplicity, but also 
because the pitch angle scattering is likely to be an efficient
process during the reacceleration {\it phase}.
 We have carried out several tests and found that, 
due to the low magnetic field intensities, 
the differences between the synchrotron spectrum 
(and all the other relevant quantities 
in this paper) calculated in the isotropic and
anisotropic case are negligible.

\subsection{Matching the observations: 
the inferred B field distribution}

From Eqs.(\ref{sincroemiss}--\ref{alpha}), at each $r$ and for
a given set of the parameters ($\chi_{LS}$, $\chi_{G0}$, 
and $B$), we obtain the synchrotron 
emissivity as a function of the frequency measured in terms 
of the break frequency $\nu_b(r)$
($\nu_b(r) \propto B(r)\gamma_b^2(r; \theta=90^o)\,$)
and then we can compute the 
spectral index between 327 MHz and 1.4 GHz.

In Fig.7 we plot the calculated intrinsic 
spectral index between 327 MHz and 1.4 GHz 
at $r=5$ arcmin as a function of $B$ and for
different values of the reacceleration
efficiency $\chi$. 
\begin{figure}
\resizebox{\hsize}{!}{\includegraphics{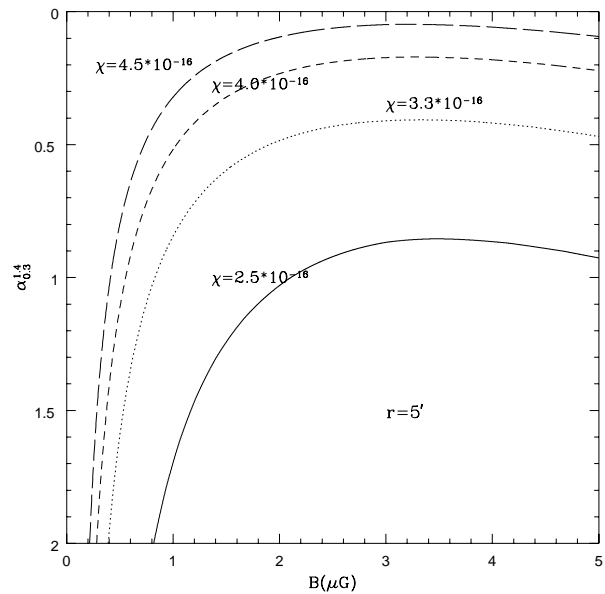}}
\caption[]{
The calculated 0.327--1.4 GHz spectral index is given
as a function of the magnetic field $B$
at a distance $r=5$ arcmin.
The calculations are performed for several 
reacceleration parameters $\chi$, as indicated in 
the panel, and for the value of $\xi$   
obtained from Eq.(\ref{thermdis}). The reacceleration
time interval is $\tau$ = 0.8 Gyr.
}
\end{figure}

\noindent
Small values of $B$ lead to very steep radio spectra
because the break frequency
of the synchrotron spectrum 
is comparable or smaller than 327 MHz.
Moderate values of the magnetic field allow to obtain
very flat spectra as those derived for the
central regions of the Coma halo ($\alpha_{0.3}^{1.4}\sim 0.4-0.5$ 
at $r=5^{\prime}$).
Since from simple synchrotron relationships
 the break frequency is 
$\nu_b\propto B (B_{IC}^2 + B^2)^{-2}$,
it is found that
the value of $B$ for which
the spectral index is the
smallest for any given reacceleration $\chi$ is
obtained when $B \sim B_{IC} \simeq 3\mu G$,
while the value of
$B$ for which one obtains a given spectral index 
depends on $\chi$.

\begin{figure}
\resizebox{\hsize}{!}{\includegraphics{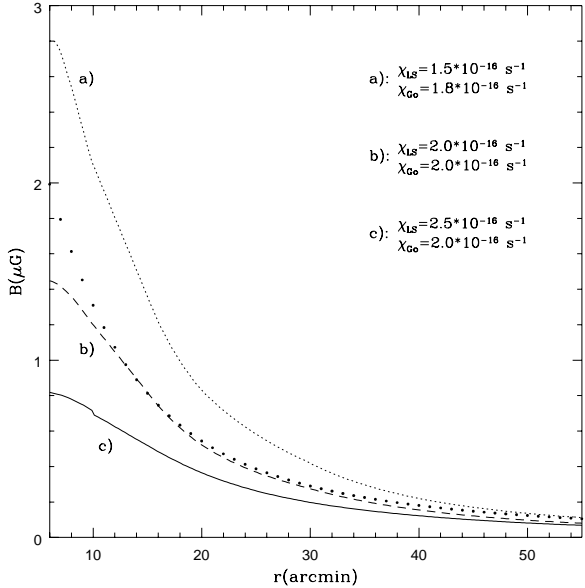}}
\caption[]{
The $B$ distribution required by the model to match
the spectral observations is reported for different
reaccelerations as indicated in the panel.
The calculations are performed for a reacceleration
period $\tau =0.8$ Gyr.
Due to the reacceleration profile (Eq.\ref{alpha})
the magnetic field in the majority of the 
cluster volume depends only on $\chi_{LS}$.
The dots represents the Jaffe (1980) theoretical 
predictions for the magnetic field distribution
in the Coma cluster (we have assumed $B_0=3\mu G$).}
\end{figure}

The magnetic
field $B(r)$ is numerically derived
by imposing that the calculated spectral index
matches the intrinsic one of Fig.1 (the average between
Deiss et al (1997) and Giovannini et al (1993) curves).
The results are shown in Fig.8 for relevant values of the 
reacceleration parameters.
For relatively small distances from the cluster center 
the value of the magnetic field is sensitive 
to the assumption about both $\chi_{LS}$ and $\chi_G$,
while at larger distances, due to the rapid decrease of the
small scale reacceleration efficiency 
(Eq.\ref{thermdis}), the value of $B$ 
depends only on $\chi_{LS}$.

With reference to Fig. 7 and $\alpha = 0.5$, we find 
a lower bound to the central value of the reacceleration 
efficiency $\chi(0) \sim 3.3\cdot 10^{-16}$ s$^{-1}$ and
a corresponding magnetic field strength 
$B(0)\sim 3 \mu G$ which is the largest value of the 
central field intensity compatible with our model. 

We emphasize
that these results depend only on the radio spectral indices
and not on the normalizations of the relativistic electron 
distributions.

The magnetic field strength 
required by our model to match the spectral indices in 
the central region can be compared with the values
given in the literature, obtained by Faraday rotation measurements. 
Kim et al. (1990) 
have estimated a magnetic field $B = (1.7 \pm 0.9) \mu G$ 
in the central
regions of the Coma cluster,
assuming a tangling scale of the magnetic field of 10-40 kpc.
More recently Feretti et al. (1995), studying the RM 
distribution of the
head-tail radio source NGC 4869 located near the center of Coma, 
inferred the existence of a strong magnetic field component
($B\sim 6 \mu G$)  tangled on the kpc scale.
The two values are not in conflict, considering that the 6 $\mu$G 
field could be associated to the local environment around NGC4869.
The value obtained by our model in the central Coma region (
$\sim 1-3 \mu G$)
is consistent with the above measurements. It is also
in agreement with the results of Clarke et al.(1999), who,
using statistical RM technique
in a sample of low redshift rich Abell clusters, 
find central values of the magnetic field of the order of 
a few $\mu$G.

The RM results can be used to constrain the reacceleration 
parameter from above.
With reference to Fig.7 we find that 
the central value of the magnetic field is 
significantly
smaller than 1 $\mu G$ for $\chi(0) > 5 \cdot 10^{-16}$ s$^{-1}$, 
setting an upper bound to the reacceleration efficiency in our
model.

The origin and 
structure of the magnetic fields in galaxy clusters
are presently debated.
Seed fields could be generated by  
turbulent galactic wakes (Jaffe 1980) and/or
by ejecta from galaxies (Kronberg et al. 1999; V\"{o}lk \&
Atoyan 1999) and maintained by turbulent dynamo
driven by galaxy motions through the ICM (De Young 1992).
Such low level magnetic fields can also be amplified
by cluster mergers (Norman \& Bryan 1998; Eilek 1999).
According to Jaffe (1980),  
the magnetic field
distribution depends on the thermal 
gas density ($n$) and on the distribution of massive 
galaxies ($n_{G}$) as 
$B(r) = B(0) (n(r)/n(0))^{0.5}
(n_{G}(r)/n_{G}(0))^{0.4}$.
Jaffe (1980) predictions calculated for the Coma cluster, 
adopting $B(0)\sim 3 \mu G$ and with $n(r)$ from   
Briel et al.(1992) and $n_G(r)$ from Girardi et al.(1995), 
are consistent with our model findings (Fig.8).

\subsection{The predicted total synchrotron spectrum}

The model at this point is anchored to 
the observed and intrinsic 327--1400 MHz  spectral
index distribution as a function of $r$ and to
the observed FWHMs of the brightness distribution at
these frequencies.

\begin{figure}
 \resizebox{\hsize}{!}{\includegraphics{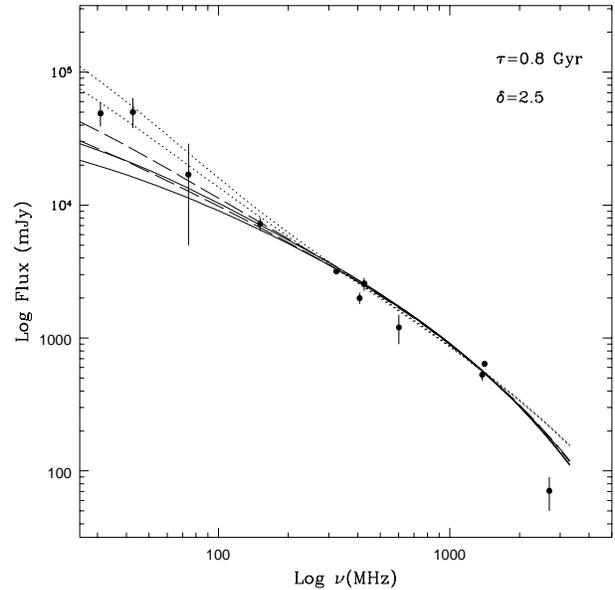}}
\caption[]{The total synchrotron radio spectrum 
predicted by
our model is compared with the flux measurements 
for the three model's parameters specified in Fig.8: model a)
(dotted lines), model b)(dashed lines) and model c)(solid lines).
The calculations are given for two different values of 
$f$ (Eq.\ref{alpha}): 0.01 (lower lines), 0.02 (upper lines).}
\end{figure}
The spectral shape of the synchrotron emission
at a given radius is very different from a simple
relatively steep power law such as the total
Coma C radio spectrum. The shape of the electron energy
distribution changes with $r$ in a predictable way depending
on losses and gains which, as shown in Section 4.2, are now 
well constrained.
 
As a consequence, a fundamental test of the model is 
whether it can
reproduce the observed total synchrotron spectrum.
This is obtained by
integrating the synchrotron emissivity (Eq.10)
over the cluster volume
taking into account the $r$--dependence of the
parameters involved in the calculation.
It is at this point that the electron spectra are 
normalized as a function of $r$ in
order to match the {\it observed} emissivities (Eq.14) derived from 
the brightness distributions at 327 and 1400 MHz following
Deiss et al (1997) procedure.
As an illustrative example
in Fig.10 we plot the electron energy distribution
at different distances from the cluster center 
for the representative case c) of Fig.8.

\begin{figure}
 \resizebox{\hsize}{!}{\includegraphics{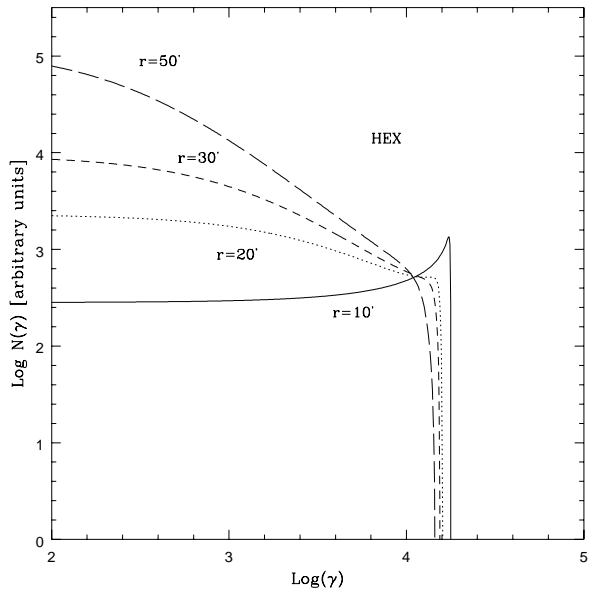}}
\caption[]{
The electron energy spectra are 
reported at different distances  
from the cluster center (normalization scale in arbitrary units).
The calculations are performed for 
model c) with $f$=0.015 (Fig.8).
The energy interval of the electrons emitting hard X--rays
via IC scattering of the CMB photons is also 
shown (HEX).}
\end{figure}

The calculated total synchrotron spectra for the same 
set of parameters of Fig.8 are represented
in Fig.9 and compared with the observations.
Once the uncertainties 
in the observations are considered,
the model well accounts for the observed 
radio spectrum over all the frequency range
sampled by the observations. 
Since the model is mainly forced on the 
327 and 1400 MHz observations, it is not
surprising that the computed spectra 
between these two frequencies are essentially 
indistinct.
On the contrary, at higher and  
lower frequencies the calculated synchrotron
spectrum is more sensitive to the assumed parameters.
In order to see whether this effect can help in
further constraining the model parameters it is important
to assess the quality of the spectral measurements.

The radio fluxes quoted in the
literature are obtained by subtracting from the 
observed fluxes those contributed by  
point sources falling within the halo.
Since the observations at different frequencies
have been performed with very different instruments and,
in some cases, at times differing of many years, the
subtraction of the sources is not standard and 
can influence the calculation of the halo flux.
In particular, the low frequency points at 43 and 74 MHz
in Fig.9 were obtained by subtracting the
extrapolated flux from 5C4.81 and 5C4.85 and a 
contributed flux from the unresolved sources measured at 
430 MHz and extrapolated down to low frequencies 
(Hanisch \& Erickson 1980); 
the 30.9 MHz halo flux 
has been obtained by subtracting all the sources 
from the 151 MHz survey with an extrapolated 
flux density at 30.9 MHz larger than 1 Jy (Hanning 1989). 
More precise subtraction procedures 
have been applied at higher frequencies 
thanks to the better spatial resolution:
the 327 MHz radio flux of the Coma halo
has been obtained by subtracting the 327 MHz flux densities  
from a list of 64 unresolved sources observed at 
327 and/or 1400 MHz (Venturi et al. 1990) and
a similar procedure has been adopted by Deiss et al.(1997)
at 1.4 GHz by using the list of radio sources as compiled by 
Kim et al.(1994) (the 2.7 GHz flux is likely to be 
underestimated (Deiss et al. 1997) ).
We find that the detailed high frequency subtractions are
deeper than those performed at lower frequencies
by 20--40 \%, so that 
the 30.9, 43, and 74 MHz fluxes reported
in Fig.9 may be considered as upper limits. 
It follows
that the low frequency data cannot discriminate among the
the models of Fig.9.

In our model 
the radio spectrum at low frequencies ($\nu < 100$ MHz)
is dominated by the contribution of the radiation emitted 
at large radii ($r > 12-17$ arcmin).
The contribution from the particle with
very low break frequencies at $r > 30-35$ arcmin,  
not detected in the
high frequency maps due to ageing, is also important. 
Larger values of $f$, corresponding to an increased number of
particles at large distances from the center, do not 
appreciably modify the 327--1400 MHz spectrum,  but
increase the low frequency 
radio fluxes expected by the model (Fig.9).

By increasing the large scale reacceleration 
($\chi_{LS}$), the spectrum of the synchrotron electrons 
is depressed and the expected radio fluxes at 
low frequencies decrease (Fig.9).
On the other hand, by decreasing $\chi_{LS}$
the low frequency radio spectrum steepens and may significantly
overproduce the observations (Fig.9).
This allows us to
obtain a range for the large scale reacceleration
efficiency of $1.5 \cdot 10^{-16}$s$^{-1}$ 
$< \chi_{LS} < 3 \cdot 10^{-16}$s$^{-1}$.
Accurate measurements at low frequencies, such as those
obtainable with the VLA at 74 MHz, are of 
crucial importance to constrain the model parameters.
 
Due to the very steep radio spectrum of the external
regions, the high frequency spectrum of the Coma halo
does not depend on $f$ (Fig.9).
It is emitted by regions at $r < 12$ arcmin
and depends on both the large scale and small scale  
reacceleration efficiencies.
With increasing reacceleration in the cluster core 
the particle spectrum flattens and 
the break energy increases: this causes a decrease
of the required $B$ value so that
the 327 MHz emitted frequency is closer to
the synchrotron break frequency.
As a net consequence
the high frequency radio spectrum steepens (Fig.9).

\subsection{Relic electrons and halo energetics}

\begin{figure}
 \resizebox{\hsize}{!}{\includegraphics{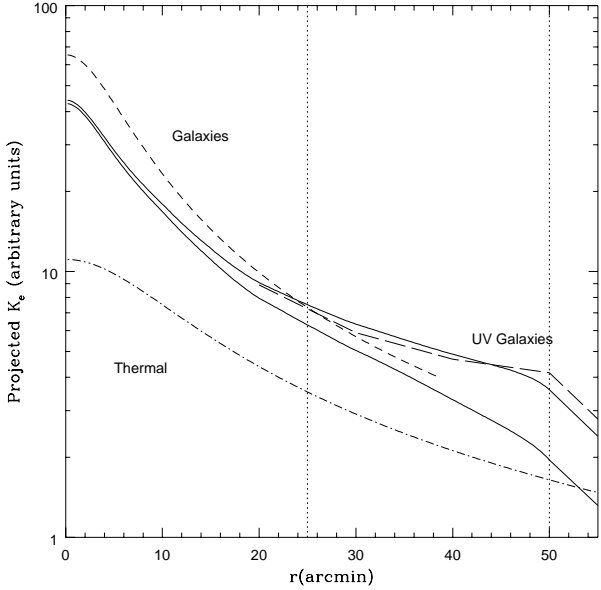}}
\caption[]{
The projected distribution of the injection rate of the 
relativistic electrons $K_e(r)$ (Eq.\ref{fase1})
during the {\it first phase} is given (solid lines)
as a function of $r$ (in arcmin) for model (c) of Fig.8
($f$= 0.01 and 0.02 starting from the bottom).
The shapes of
the projected density of the thermal gas (obtained from 
Briel et al.1992 fit) 
and of the galaxies (Girardi et al.1995) 
(dashed lines) are conveniently scaled for a closer comparison;
the projected galaxy density profile is only valid up to 38 arcmin 
from the cluster center.
The projected radial distribution of the UV galaxies
(Donas et al.1995), normalized to the Girardi et al. counts,
is also plotted for $r > 25$ arcmin.
The region in which the 80-90 \% of the 
IC hard X--ray flux is produced during the {\it
second phase} (Section 5) is enclosed by the
vertical dotted lines.}
\end{figure}
 
The shape of the injection rate (projected) of the
relic electron population, $K_e(r)$, is shown in Fig.11. 
It has been   
obtained by inverting the synchrotron emission
equations for model (c) of Fig.8
(the integral of Eq.\ref{sincroemiss} 
along the line of sight).
The shape is rather close, within a factor $\sim 2$,  
to both the thermal and galaxy number density distributions
in Coma. This result is very encouraging since no quantitative
assumptions on the space distribution of the relic electrons
have been made in the derivation of the model. However, the 
similarity with the matter distribution is consistent with our
conjecture that relativistic particles must have been 
efficiently injected into the ICM during the {\it first phase} 
by AGN and galaxy activity.

The moderate upturn of $K_e$ at $r > 45 $ arcmin
is mainly caused by the flattening in the
assumed Coma radio brightness due to the superposition 
of the two Gaussian functions (Eq.\ref{radiobr}). In any case
this upturn is not relevant for the main characteristics
and predictions of the model.

The total energy associated with the relativistic electrons
in the cluster is the result of the 
energy injected during the {\it first phase} and
of the energy gained by the electrons during
the {\it second phase}.
We find that for a reacceleration parameter 
$\chi \sim 3 \cdot 10^{-16}$s$^{-1}$ 
the total energy in  
relativistic electrons is $\sim 10^{59}$erg.

We point out that this value
is much smaller (by a factor $>10$) 
than that associated with the relativistic
electron population considered in the other
proposed models (Kempner \& Sarazin 1999 and
references therein). This is because the rather 
flat electron energy spectrum, characteristic 
of our model, is much more efficient than those
of the other models in emitting synchrotron 
radiation at radio wavelengths.

\section{The hard X--ray tail in the Coma cluster}

From a deep BeppoSAX observation  
Fusco--Femiano et al.(1999) discovered 
an hard X--ray tail, in the range 20--80 keV, 
exceeding the extrapolation 
of the thermal X-ray emission spectrum; 
this excess has also been confirmed by an RXTE 
observation (Rephaeli et al. 1999) with consistent
results.
Due to the poor statistics, the spectrum is not well constrained
so that the origin of this high 
energy emission cannot be firmly established. 
Several models
have been explored.
It can be generated 
by the IC scattering of the radio emitting electrons with the 
CMB photons (Fusco--Femiano et al. 1999; Sarazin et al. 1999; 
Sarazin \& Lieu 1998); in this case models invoking a secondary 
production of the electrons yield a gamma--ray flux considerably  
larger than the EGRET upper limit (Blasi \& Colafrancesco 1999).
%If this is the case, then in the halo 
Alternatively, 
it might originate via relativistic 
bremsstrahlung of a supra--thermal power 
law tail of particles (Ensslin et al. 1999), or,
as proposed by Dogiel (2000), emitting  
from the thermal emission of
a modified maxwellian distribution of the hot ICM
induced by strong acceleration 
processes ($\chi \sim 10^{-15}$s$^{-1}$). 

The IC origin is particularly attractive in view of
the fact that a comparison between 
the radio emission from the halo and the hard X--ray tail
discovered by BeppoSAX, if spatially coincident,
may allow a direct estimate of
the average strength of the magnetic field $B$ over the 
cluster volume. 
Nevertheless, since in the present model 
%the break energy of 
the electron energy spectrum
depends on the distance from the cluster center (Fig.10),
the synchrotron as well as the IC spectra are emitted
by relativistic electrons whose
energy distribution cannot be represented by a unique power law 
over the whole volume.
Therefore, the magnetic field cannot be simply evaluated by
standard formulae (e.g. Harris \& Grindlay 1979).

Under our assumptions (Eq.\ref{fase2})  
the IC emissivity per unit energy and solid angle 
in the ultra--relativistic Thompson approximation 
is given by:

\begin{eqnarray}
j(\epsilon_1,r)=
{{2 r_0^2 \pi^2 }\over{ c^2 h^3}}
\int \int_0^{\pi} {{ 
d\epsilon d\theta sin \theta
}\over{ e^{\epsilon/ k T_{CMB}} -1}}
\int_{\gamma_{min}}^{\gamma_b(r,\tau,\theta)}
{{d\gamma}\over{ \gamma^{\delta+3} }}
\nonumber\\
\left(1- {{\gamma}\over{\gamma_b(r,\tau,\theta)}} 
\right)^{\delta-3}
\left( 1+ {{\xi(r) \gamma^{-1} -\chi(r)}\over
{\gamma_b(r,\tau,\theta) \beta}} \right)^{-\delta+1} \cdot 
\nonumber\\ 
\{ \gamma^2
[ {{1 + (\xi(r) \gamma^{-1} -\chi(r))(\gamma_b(r,\tau,\theta)
\beta)^{-1} }\over{
1- \gamma/\gamma_b(r,\tau,\theta) }} ]^2 \nonumber\\
+ {{\xi(r)}\over{\beta}} \}^{-1}
{{K_e(r)}\over{\delta-1}}
{\cal C}(r,\theta) 
K_{IC}[{{\epsilon_1}\over{\epsilon}}, \gamma ]
\label{ictot}
\end{eqnarray} 

\noindent
where 

\begin{equation}
{\cal C}(r,\theta) =  q(r,\theta) 
{{  (1-\tanh^2(\tau\sqrt{q}/2)) }\over{
[2 \gamma_b(r,\theta,\tau) 
\tanh (\tau\sqrt{q}/2)]^{2} }}  \beta^{-3}(r,\theta)
\end{equation}

\begin{equation}
K_{IC}\left[{{\epsilon_1}\over{\epsilon}}, \gamma
\right] =
\epsilon_1 
\left( 2\epsilon_1 ln {{\epsilon_1}\over{4\gamma^2 \epsilon}}
+\epsilon_1 +4\gamma^2 \epsilon 
-{{\epsilon_1^2}\over{2\gamma^2 \epsilon}} \right)
\label{ickernel}
\end{equation}

\noindent
$\epsilon$ being the energy of the CMB photons and
for ultra--relativistic electrons 
$\gamma_{min} = \sqrt{\epsilon_1/ 4 \epsilon}$ 
(e.g. Blumenthal \& Gould 1970). 
As a cluster volume we assume a sphere of 
radius 55 arcmin ($\sim$2.2 Mpc), that 
is about the radius up to which the counts of UV 
galaxies, presumed  
cluster members, begin merging the background
galaxies (Donas et al. 1995).

Since the IC emission is only sensitive to  
the number of electrons, 
it is expected to be mainly contributed  
by the outer parts $r>30$ of the cluster volume  
which contain
the large majority of the relativistic electrons
able to IC scatter the CMB photons in
the BeppoSAX hard band (Fig.10). 
As a consequence, the predicted X--ray luminosity is 
sensitive to the large scale reacceleration efficiency;
we find that the from 80 to 90\% of the 
model 20--80 keV flux is emitted between 30--50 arcmin
from the center (Fig.11).
 
\begin{figure}
 \resizebox{\hsize}{!}{\includegraphics{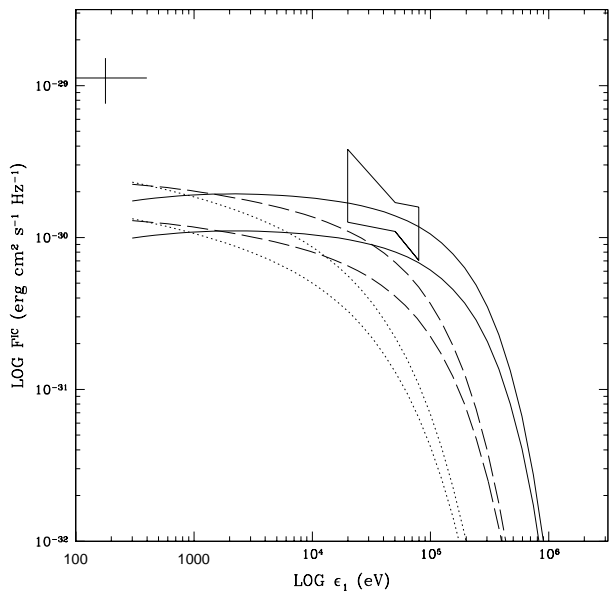}}
\caption[]{
The IC fluxes
from the Coma cluster for the three models of Fig.8
(lower line $f$=0.01, upper line $f$=0.02). 
The BeppoSAX hard X--ray data are obtained from 
Fusco--Femiano et al. (1999) with the 
spectral indices at the 90\% confidence level; a $15\%$ 
uncertainty in the normalization is assumed.
The cross represents the Coma EUV excess Bowyer et al.(1999).}
\end{figure}

The calculated hard X--ray IC fluxes from our model
for the three choices of the parameters given
in Fig.8
are compared with the BeppoSAX data in Fig.12.  
Models with relatively high reacceleration 
efficiencies may account for a large fraction or all of
the X--ray flux measured by BeppoSAX. By comparing Fig.12
with Fig.9 it is seen that it is precisely this type of
models that contribute less to the synchrotron flux
at very low radio frequencies. It is not surprising, then, 
that in order not the overproduce the BeppoSAX
flux we obtain an upper limit for the large scale
reacceleration efficiency $\chi_{LS} < 3\cdot
10^{-16}$s$^{-1}$, the same as that found in Section 4.6.
Obviously, accurate measurements at low radio frequencies
would be of crucial importance not only to constrain the 
parameters of our model for Coma C, but also to check whether
an IC origin of BeppoSAX hard X--rays is viable.
Thus, at variance with statements found in the literature, 
central magnetic fields of $\sim 1\mu G$ 
may be consistent with an IC origin of the
BeppoSAX hard X--ray flux, the reason being that while
the X--rays are mainly produced in the outer regions
of the cluster the radio flux above a few 100 MHz is mainly
produced in the inner regions. 

In Fig.12 we have also represented 
the EUV flux from Bowyer et al.(1999):
the model EUV flux is well below the observed.
The origin of the EUV excess is still unclear, its
spatial distribution is considerably narrower than the
radio brightness distribution and, if non--thermal,  
it is probably associated to an additional 
electron population (Bowyer \& Bergh\"{o}fer 1998;
Ensslin et al. 1999; Brunetti et al. 1999).

\section{Discussion and conclusions}

\subsection{A two phase model for Coma C} 

It has been shown that the radio properties of
Coma C can be explained by a {\it two phase} model
consisting of an injection period, in which a power 
law spectrum of relativistic electrons has been 
continuously injected throughout the
cluster volume and modified by radiative and Coulomb 
losses, followed by a reacceleration phase of the aged 
(relic) spectrum for
typical duration of $\sim$1 Gyr up to the present time.

The model is anchored to the 
spectral index map of Coma C between 327 and 1400 MHz
which shows a marked steepening going in the outward 
direction from a flat central value. The explanation of
this pattern, not accounted for in other published models,
constitutes the basic motivation of this work.

On the basis of simple considerations (Section 2.3)
we have shown that the reacceleration efficiency should be
$\chi \sim 2\cdot 10^{-16}$s$^{-1}$ and the reacceleration time
$\tau > 0.6$ Gyr. A detailed modelling has then been worked out
based on the following assumptions:

$\bullet$
The brightness distributions at 327 and 1400 MHz are 
described as the sum of two gaussians; the radio images constrain 
the peak amplitude of the large scale gaussian to be a fraction 
$f < 0.06$ of that of the small scale gaussian.   
By assuming a spherical symmetry,
the emissivities at 327 and 1400 MHz and the corresponding
spectral index as a function
of the radial distance $r$ are obtained from the observed 
brightness distributions 
by making use of Deiss et al. (1997)
procedure. 

$\bullet$ 
The Coulomb losses, $\xi (r)$, are fixed by the
$\beta$-model fit of the ICM of Briel et al (1992).

$\bullet$ 
The reacceleration efficiency is assumed to be 
represented by the sum of two terms: a large scale reacceleration
($\chi_{LS}$) constant throughout the cluster volume and a small
scale reacceleration ($\chi_G (r)$) peaking at the cluster center.
We have identified the large scale component with 
shocks and/or turbulence,  
possibly generated during a recent merger, and the
small scale component with the amplification of
such shocks/turbulence by the motion of the
massive galaxies in the cluster core.

$\bullet$ 
Let evolve the relic electron population, subject to 
radiation and Coulomb losses, for a reacceleration time 
$\tau = 0.8$ Gyr.

With these assumptions we find that:

$\bullet$ 
The electron energy distribution depends on $r$, being
much flatter and with a somewhat higher energy break
toward the cluster center where both the magnetic field 
intensity and the reacceleration parameter are stronger
and the Coulomb losses are more severe.

$\bullet$ 
The value of the 327--1400 MHz spectral index in the 
central region ($\alpha = 0.5$) implies an upper bound to the 
intensity of the magnetic field $B \sim 3\mu G$ and a lower
bound to the reacceleration efficiency $\chi(0) \sim 3.3\cdot 10^{-16}$
s$^{-1}$.

$\bullet$ 
A central magnetic field of at least 1 $\mu G$, as 
indicated by RM measurements, implies an upper bound to the
reacceleration efficiency at the center of $\chi(0) \sim 5\cdot
10^{-16}$s$^{-1}$.

$\bullet$ 
By fitting the steepening of the 327-1400 MHz spectral
index with $r$, the magnetic field intensity smoothly tapers off
from central values of 1--3 $\mu G$ to $\sim 0.1 \mu G$ at 
$r \sim 50$ arcmin.

$\bullet$ 
The model synchrotron emission perfectly fits the total
radio spectrum of Coma C within 327 and 1400 MHz, independent of
the $\chi$ and $f$ parameters allowed to vary within the limits
set above. This demonstrates the internal consistency of the model.
On the contrary, the computed low frequency fluxes vary appreciably
with the parameters, being higher with decreasing $\chi$ and 
increasing $f$: most of this flux is emitted by the outer parts of
the cluster volume, so that by a closer comparison with the observed
low frequency flux it is possible to set limits on the large
scale acceleration efficiency $1.5\cdot 10^{-16} < \chi_LS < 3\cdot
10^{-16} s^{-1}$.

$\bullet$
With the electron energy and space distributions constrained by the 
radio data it is found that the X-rays produced by IC
scattering the CMB photons can account for the hard X-ray flux 
discovered by BeppoSAX if the larger figures of the $\chi_LS$ and $f$
parameters, allowed by the radio model, apply. Like the flux
emitted at low radio frequencies, most of the X-rays are produced 
in the outer volume of the cluster. This demonstrates once again
the crucial importance of accurate flux measurements at frequencies
$\leq$ 100 MHz.
We remark that the BeppoSAX flux is significantly
overproduced for $\chi_{LS} > 3\cdot 10^{-16}$s$^{-1}$
($f=0.02$), which is consistent with the upper limit 
set by the radio spectrum at low frequencies.

$\bullet$
By de-evolving the electron spectra constrained by the radio model
it is found that the shape of projected radial distribution of the 
injection rate of the relic electron population is very similar to 
that of the galaxies in the Coma cluster. This is consistent with
our conjecture that the relic electron population may have been
injected by AGN and galaxy activity at earlier epochs.
 
$\bullet$
The present energy in relativistic electrons required by our
model is $\sim 10^{59}$ ergs, at least 
a factor 10 smaller than in
other proposed models.  
The energy budget required at the end of the injection phase is
$\sim 2\cdot 10^{58}$ ergs, if almost immediately followed by the
reacceleration phase, and becomes larger by increasing the time gap 
between the two phases simply because the spectrum
of the relic electrons is further depressed. We find that a temporal
gap of 3--4 Gyr involves an energy budget 
$\sim$ 100 times larger, but still
reasonable (see Appendix).
           
We emphasize that the results listed above are essentially obtained 
by imposing
that the model accounts for the radio properties of Coma C and 
we find that 
this requirement is already stringent enough to constrain the relevant
parameters, such as the reacceleration's efficiencies and time, within
a narrow range. We also notice that the typical reacceleration 
efficiency, corresponding to an e-folding time scale of 
$\sim$ 0.1 Gyr, is relatively modest and 
comparable or smaller than the values
frequently quoted both in clusters (cluster weathering) and in tailed
radio galaxies.

Finally, the likely presence of relativistic protons injected
with the electrons deserves some attention. It  
can be constrained by looking for the gamma--ray emission 
from the decay of the $\pi^0$ generated in the
p--p collisions.
Recently, Blasi \& Colafrancesco (1999) have worked
out a secondary electron model for Coma C and 
estimated an upper bound $\sim 4\cdot 10^{63}$erg
($H_0=50$ km s$^{-1}$ Mpc$^{-1}$) for the
total energy of centrally injected protons 
in order not to exceed 
the upper limit on the gamma--ray flux set 
by EGRET (this figure should be much larger 
in our model where the majority 
of the relativistic particles are injected in regions of low 
ICM density).
Since in our model 
the particle reacceleration is due to
shocks and/or turbulence in the ICM,  
the energy gained by relativistic protons
during the {\it second phase}
can not exceed the thermal energy of the
ICM ($\sim 10^{63}$erg) so that 
the gamma--ray flux from p--p 
collisions in our model is expected 
to be far below the EGRET limit.

\subsection{General considerations for the radio
haloes}

Although the present work is focussed on 
Coma C, we believe that a {\it two phase model} 
may explain the radio properties of other haloes
as well.

Like for Coma the required energy input in the ICM
could be supplied by a recent merger as suggested
by the evidence of a significant association
between radio haloes and cluster mergers.
Moreover, it has been found that the
occurrence of radio haloes in clusters
increases with the cluster X--ray luminosity
$L_X$ in the 0.1--2.4 keV band 
(from $\sim 4\%$ for 
$L_X < 5\cdot 10^{44}$erg s$^{-1}$ to $\sim 36\%$ for
$L_X> 10^{45}$erg s$^{-1}$;
Giovannini 1999 in preparation).
This is consistent with the expectations 
of the {\it two phase} model
since more X--ray luminous, massive clusters
would have produced a larger number of injected
particles during the {\it first phase}, 
thus supplying sufficient energy for the formation
of the radio haloes also after a moderately long
gap between the injection and the reacceleration
phases. 

It should be born in mind, however, that the strength
of the halo radio emission will critically depend 
on the combination of the various parameters,
(reacceleration efficiency and duration,
magnetic field strength and the temporal
gap between the two phases) 
so that
we expect that only a fraction of the
massive clusters with a recent merger will show
a substantial radio halo.
For instance, contrary to expectation
a too efficient reacceleration
may result in an electron spectrum very much
stretched toward higher energies so that, other
conditions being equal, 
there may be fewer relativistic electrons
emitting at radio frequencies in a given 
magnetic field and the halo emission would be
dimmed.

The IC emission of the relativistic electrons
with the CMB photons will typically fall in the
UV and EUV band at the end of the injection phase, 
progressively moving toward
the X--ray band during the reacceleration
phase.
Thus, non--thermal X--ray emission from clusters
with recent mergers should be a quite
common feature, depending on the
total number of relic electrons.
In general, as seen in the case of Coma,
the radio at intermediate frequencies and the hard X--rays 
are expected to be mainly contributed by 
different regions : the former near the center 
of the cluster, associated with 
regions of relatively higher $B$, the latter  
 from the outer parts of 
the cluster volume.
Clearly, observations with future
hard X--ray facilities  
will be of great value in checking
the correctness of this scenario.

\section{Acknowledgements}

We are grateful to A.Atoyan, S.Colafrancesco, and
T.Ensslin    
for useful discussions, and to M.Murgia for providing a 
numerical table of the synchrotron electron Kernel.
We wish to thank the referee whose criticism has stimulated
us to work out an improved version
of the paper.
This work was partly supported by the Italian Ministry for
University and Research (MURST) under grant Cofin98-02-32, 
and by the Italian Space Agency (ASI).

\appendix{}
\section{The general two phase model}

In Section 3 we have analyzed the effect due to the 
systematic reacceleration
on the energy distribution.
In order to simplify the scenario, it was 
assumed that the reacceleration phase 
starts when the injection phase 
is stopped or strongly reduced. 

In this Appendix we generalize the results of this 
paper by assuming that 
the two phases are separated in time.

After the injection is stopped at a given time $t_i$,  
we let the particles 
to age for a time interval $\Delta t$.
Mostly due to IC losses with the CMB photons, 
the particles rapidly age and the energy distribution
is depopulated at relatively high energies.
%for $\gamma > 10^3$.

By integrating Eq.(\ref{fase1losses}) with constant losses
one obtains:

\begin{equation}
\gamma(\theta,\Delta t) =
{{ \gamma(t_i) - \sqrt{\xi/\beta(\theta)}
\tan ( \sqrt{\xi \beta(\theta)} \Delta t) }\over
{1+ \gamma(t_i) \sqrt{\beta(\theta)/\xi} 
\tan ( \sqrt{\xi \beta(\theta)} \Delta t) }}
\label{a1gammaint}
\end{equation}

\noindent
and a break energy :

\begin{equation}
\gamma_{b,1}(\theta) =
{1\over{ \sqrt{\beta / \xi} 
\tan( \sqrt{\xi \beta} \Delta t) }} \,
%{ \buildrel \xi \rightarrow 0  \over 
%\longrightarrow }
%\,
%{1 \over{ \Delta t \beta(\theta)}}
\label{a1gammabint}
\end{equation}

\noindent
Then, 
from the kinetic equation (Eq.\ref{kinetic}; with $Q=0$), 
one finds the energy distribution of the electrons :

\begin{eqnarray}
N(\gamma,\Delta t,\theta) =
{{K_e [1+ \tan^2 (\sqrt{\beta \xi} \Delta t)]}\over
{\beta(\theta) (\delta-1)}}
\left(1- {{\gamma}\over{ \gamma_{b,1} }} 
\right)^{\delta-1}
\cdot \nonumber\\
\left(\gamma + {{\xi}\over{\beta(\theta) 
\gamma_{b,1} }} \right)^{-(\delta+1)}
\left( 1 + 
( {{1-\gamma/\gamma_{b,1} }\over
{ \gamma + {{\xi}\over{\beta(\theta) 
\gamma_{b,1} }} }} )^2
{{\xi}\over{\beta(\theta)}} \right)^{-1}
\label{a1numint}
\end{eqnarray}

\noindent
which becomes Eq.(\ref{fase1}) for $\Delta t \rightarrow 0$.

We assume that after a time  
$\Delta t$ following the end of the injection phase, 
the particles are reaccelerated.
The time evolution of the energy of the
electrons is obtained by integrating Eq.(\ref{fase2losses}):
%\begin{equation}
%{{d\gamma}\over{dt}}
%=-\beta(\theta) \gamma^2 -\xi+\gamma \chi
%\label{a1fase2losses}
%\end{equation}
%\noindent
%that integrated yields 

\begin{eqnarray}
\tau =
\cases{q<0 \, \Rightarrow \cr
{2\over{ \sqrt{q} }} \left(
\tan^{-1} {{ \chi -2\xi \gamma(\tau)}\over
{ \sqrt{q} }}  
-
\tan^{-1} {{ \chi -2\xi \gamma(\Delta t)}\over
{ \sqrt{q} }} \right)  \cr
q>0\, \Rightarrow \cr
{-2\over{ \sqrt{-q} }} \left(
\tanh^{-1} {{ \chi -2\xi \gamma(\tau)}\over
{ \sqrt{q} }} 
-
\tanh^{-1} {{ \chi -2\xi \gamma(\Delta t)}\over
{ \sqrt{q} }} \right) \cr }
\label{a1integral}
\end{eqnarray}

\noindent
where
$q(\theta) = \chi^2-4\xi \beta(\theta)$ 
expresses the importance of the reacceleration.

\noindent
Eq.(\ref{a1integral}) can be 
explicated in terms of $\gamma$; it gives :

\begin{equation}
\gamma(\tau,\theta)=
{{\sqrt{-q}}\over{2 \xi}}
{{ 2\xi\gamma(\Delta t) -\chi -
\sqrt{-q} \tan (x) }\over
{ (2\xi\gamma(\Delta t) -\chi) 
\tan (x) -\sqrt{-q} }}
+ {{\chi}\over{2\xi}}
\label{a1gamma+}
\end{equation}

\noindent
for $q<0$, while

\begin{equation}
\gamma(\tau,\theta)=
{{ \gamma(\Delta t) 
(\sqrt{q}/ \tanh(x) + \chi ) -2\xi }\over{
2\beta(\theta) \gamma(\Delta t) + \sqrt{q} / \tanh(x) -\chi}}
\label{a1gamma-}
\end{equation}

\noindent 
for $q>0$, 

\begin{equation}
x =
{{\sqrt{|q|}}\over 2}
\tau 
\label{a1time}
\end{equation}

\noindent
and $\tau$ is the reacceleration period.
Eq.(\ref{a1gamma-}) reduces to 
Eq.(\ref{a1gammaint}) for $\tau=0$.

\noindent
The break energies are obtained from Eqs.(\ref{a1gamma+} \&
\ref{a1gamma-})
by imposing $\gamma(\Delta t) \rightarrow \infty$ :

\begin{eqnarray}
\gamma_b(\tau,\theta)=
\cases{
q<0 \Rightarrow \cr
\sqrt{-q} \{ 2\xi 
\tan ( x ) \}^{-1}
+ \chi / 2\xi  \cr
q>0 \Rightarrow \cr
\{ \chi
+ \sqrt{q}/ \tanh(x) \}
( 2\beta(\theta))^{-1} 
\cr }
\label{a1gammab}
\end{eqnarray}

\noindent
In the case of the Coma cluster, 
$q>0$; this is because 
by assuming a physically sound reacceleration time 
$\sim 10^{8}$yr it is $\chi^2 \sim 10^{-31}$, while,  
given the thermal gas density and
the IC losses, it is $4 \xi \beta < 4\times 10^{-34}$.
The energy distribution of the reaccelerated 
electrons is obtained by solving the kinetic equation
(Eq.\ref{kinetic}, with $Q=0$) with  
Eqs.(\ref{a1numint}, \ref{a1gamma-}, \&
\ref{a1gammab}):

\begin{eqnarray}
N(\gamma,\tau,\theta) = 
{{q(\theta) K_e }\over{4}} \,
{{ 1 - \tanh^{-2} (x) }\over {\beta^3(\theta) \gamma_b^2(\theta) }} \,
{{ 1+ \tan^2 (\sqrt{\beta(\theta) \xi} \Delta t) }\over{
\delta -1 }}  \nonumber\\
\left(1- {{\gamma {\cal L(\theta)} }\over{ \gamma_{b,1} }} 
\right)^{\delta-1}
\left( 1 + ( {{1- \gamma {\cal L(\theta)} / \gamma_{b,1} }\over
{\gamma {\cal L(\theta)} + 
{{\xi}\over{\beta(\theta) \gamma_{b,1}}} }} )^2
{{\xi}\over{\beta(\theta)}} \right)^{-1} 
\cdot \nonumber\\
\left(\gamma {\cal L(\theta)} + 
{{\xi}\over{\beta(\theta) \gamma_{b,1} }} 
\right)^{-(\delta+1)} 
\left(1- \gamma / \gamma_b(\theta) \right)^{-2}
\label{a1num2}
\end{eqnarray}

\noindent
where

\begin{equation}
{\cal L(\theta)} =
\left( 1 + {{\xi \gamma^{-1} - \chi }\over{\gamma_b(\theta) 
\beta(\theta)}} \right)
\left(1 -{{\gamma}\over{\gamma_b(\theta)}} \right)^{-1}
\label{a1l}
\end{equation}
 
\begin{figure}
\resizebox{\hsize}{!}{\includegraphics{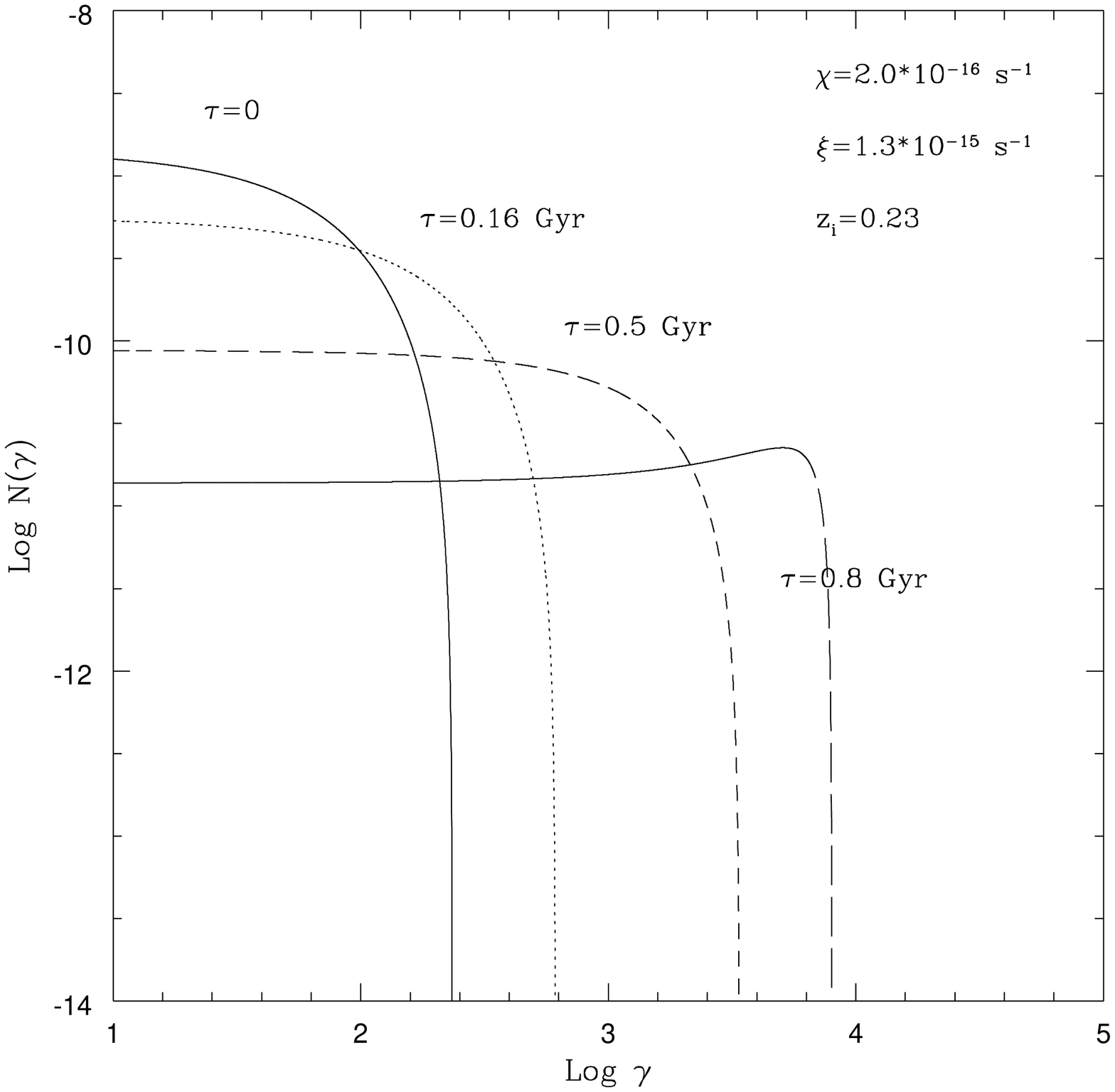}}
\caption[]{
The predicted electron 
energy distribution, in the same arbitrary units as
in Fig.3, 
is shown as a function of the reacceleration period
$\tau $ as indicated in the panel.
The calculations are performed from Eq.(\ref{a1num2}) for 
$\chi=2.0 \cdot 10^{-16}$
s$^{-1}$, $\xi = 1.3\cdot 10^{-15}$ and $z_i=0.23$.
The parameters $\Delta$ and  $\gamma_{b,1}$ can be
calculated respectively 
from Eqs.(\ref{a1deltat}) and (\ref{a1gb1}); one has :
3.4 Gyr, 230 (solid line), 3.24 Gyr, 238 (dotted line),
2.9 Gyr, 255 (short dashed line) and  2.6 Gyr, 271
(large dashed line).}
\end{figure}

The final shape of the 
energy distribution of the emitting electrons
is mostly dependent on 
the reacceleration phase during which it 
is heavily modified.
For this reason, even if the particles can be injected
and age over cosmological time scales, 
in the calculation of Eq.(\ref{a1numint}) 
we have not considered the variation
with $z$ of the IC losses.

However, the final energy distribution, Eq.(\ref{a1num2}), 
depends on $\Delta t$ and on $\gamma_{b,1}$ which  
are directly 
related to the redshift of the end of the injection
phase and to the redshift at which the reacceleration
phase starts; 
such dependences can be simply introduced in the 
model.

In the case of an Einstein--de Sitter universe, one has
that :

\begin{equation}
\Delta t = {2 \over{3 H_0}}
\left( \,
(1+z_{acc})^{-3/2}
-
(1+z_{i})^{-3/2} \, \right)
\label{a1deltat}
\end{equation}

\noindent
$z_i$ is the redshift
corresponding to the end of the injection phase, while
$z_{acc}$, the redshift corresponding to the start
of the reacceleration phase, is given by:

\begin{equation}
z_{acc}=
\left( \,
1 - {{3 H_0 \tau }\over{2 }} \, \right)^{-2/3}
-1
\label{a1zacc}
\end{equation}

\noindent
If the particles age over cosmological time
scales, the IC losses are expected to dominate
the Coulomb losses for $\gamma \geq 100$ so that
Eq.(\ref{fase1losses}) can be simplified and 
solved by considering the dependence of $\beta$ on
$z$ and that $dz/dt = H_0\sqrt{1+z}\,$; one finds:

\begin{equation}
\gamma(z_{acc})=
{{ \gamma(z_i) }\over
{ 1 + {{ \beta}\over{H_0}} b(z_i,z_{acc}) }}
\label{a1gamma}
\end{equation}

\noindent
where

\begin{equation}
b(z_i,z_{acc})=
{2\over 3} \left( \, (1+z_i)^{5/2} -
(1+z_{acc})^{5/2} \, \right)
\label{a1b}
\end{equation}

\noindent
the break energy is simply obtained from 
Eqs.(\ref{a1gamma}--\ref{a1b}) 
with $\gamma(z_i)\rightarrow \infty$:

\begin{equation}
\gamma_{b,1}=
{{3 H_0}\over{2 \beta }}
\left(
\, (1+z_i)^{5/2} -
(1+z_{acc})^{5/2} \, \right)^{-1}
\label{a1gb1}
\end{equation}

Given $\tau$ and $z_i$, from Eqs.(\ref{a1zacc}, \,
\ref{a1gb1}, \,\ref{a1deltat}) 
one has $\gamma_{b,1}$ and $\Delta t$, then from
Eq.(\ref{a1num2}) one has the particle energy
distribution at $z=0$.

In Fig.A1 the model energy distribution of relativistic
electrons injected up to a redshift $z=0.23$ is
shown for different reacceleration times.
The break of the electron energy distribution 
rapidly increases for a reacceleration time
$\tau \leq 0.8$ Gyr, then it reaches the asymptotic
value $\sim 10^4$.
During the reacceleration the IC emission 
from such a population is expected to be
channeled at the beginning in the UVs, then 
in the soft and in the hard X--rays.

By assuming that the energy 
released during a merger generates an efficient   
reacceleration of the order of $\sim 10^8$yr
in the cluster volume for $\sim$1 Gyr, then 
non--thermal hard X--ray emission should be
produced for $\sim 0.5$ Gyr.

The possibility of a gap between 
the two model phases gives a strong limit in
the detection of non--thermal emission (either synchrotron
and IC) from merging clusters.
Depending on the gap duration, 
a considerable fraction of the electrons is 
thermalized or shifted at low energies (where the
Coulomb losses dominate also during the reacceleration
phase) after the {\it injection phase}.
 
We find that, if the gap is $\sim 3-4$ Gyr long, the
energy content of the relativistic electrons 
is reduced by a factor $\sim$ 100 (Figs.3 and A1).

\end{document}